\documentclass[10pt, prb, aps, twocolumn, showpacs, citeautoscript, floatfix, reprint, amsmath, amssymb, notitlepage, superscriptaddress]{revtex4-1}

\usepackage{graphicx}
\usepackage{stmaryrd}
\usepackage{rotating}
\usepackage{amsmath}
\usepackage{amsfonts}
\usepackage{amssymb}
\usepackage{wasysym}
\usepackage[countmax]{subfloat}
\usepackage{dcolumn} 
\usepackage{bm} 
\usepackage{color}

\usepackage{float}
\usepackage{latexsym}

\begin{document}
\title{Robustness of topological order against disorder}

\author{Lucas A. Oliveira}

\affiliation{Department of Physics, PUC-Rio, 22451-900 Rio de Janeiro, Brazil}

\author{Wei Chen}

\affiliation{Department of Physics, PUC-Rio, 22451-900 Rio de Janeiro, Brazil}

\date{\rm\today}

\begin{abstract}

A universal topological marker has been proposed recently to map the topological invariants of Dirac models in any dimension and symmetry class to lattice sites. Using this topological marker, we examine the conditions under which the global topological order, represented by the spatially averaged topological marker, remains unchanged in the presence of disorder for 1D and 2D systems. We find that if an impurity corresponds to varying a nonzero matrix element of the lattice Hamiltonian, regardless the element represents hopping, chemical potential, pairing, etc, then the average topological marker is conserved. However, if there are many strong impurities and the average distance between them is shorter than a correlation length, then the average marker is no longer conserved. In addition, strong and dense impurities can be used to continuously interpolate between one topological phase and another. A number of prototype lattice models including Su-Schrieffer-Heeger model, Kitaev chain, Chern insulators, Bernevig-Hughes-Zhang model, and chiral $p$-wave superconductors are used to elaborate the ubiquity of these statements.

\end{abstract}

\maketitle

\section{Introduction}

How disorder influences the macroscopic properties of materials has been an important issue in condensed matter physics. Empirically, for materials exhibiting Landau order parameters like magnetization and superconductivity, weak disorder is usually not detrimental to the phenomena related to the order parameter in the macroscopic scale, such as the magnetic force and Meissner effect. In particular, weak disorder that respects the symmetry of the material may even have negligible influence. The most notable theory of this kind that explicitly formulates the influence of disorder is Anderson's theorem\cite{Anderson59}, which states that $s$-wave superconductivity is robust against nonmagnetic impurities that preserves time-reversal (TR) symmetry. In contrast, magnetic impurities that break the TR symmetry are detrimental to the $s$-wave superconductivity, whose presence reduces the superconducting gap\cite{Abrikosov61}.

Turning to topological insulators (TIs) and topological superconductors (TSCs), the issue of disorder becomes even more intriguing. Firstly, topological order is usually defined from the Bloch state of the filled bands in momentum space, where certain geometrical objects like Berry connection or Berry curvature momentum-integrates to a quantized integer\cite{Hasan10,Qi11}. In the presence of real space disorder, the first issue is then how to define a topological invariant given that momentum is no longer a good quantum number, especially when the disorder is relatively strong and dense. Along this line of reasoning, the notion of topological markers becomes particularly useful, which are objects derived from rewriting the momentum space topological invariant into real space via position operators and projectors to lattice eigenstates\cite{Bianco11,Prodan10,Prodan10_2,Prodan11,Loring10,Bianco13,MondragonShem14,Marrazzo17,
Cardano17,Meier18,Huang18,Huang18_2,Focassio21,Sykes21,Jezequel22,Wang22,Hannukainen22}. In particular, it is recently demonstrated that for topological materials described by Dirac models in any dimension and symmetry class\cite{Schnyder08,Kitaev09,Ryu10,Chiu16}, a universal topological marker can be introduced to map the topological invariant to lattice sites\cite{Chen23_universal_marker}, which seems to suggests that it is possible to investigate the effect of disorder in any dimension and symmetry class in a unified manner.

In this paper, we present a systematic survey on disordered topological materials within the context of the universal topological marker, which reveals a number of remarkable effects of disorder. Through investigating several prototype theoretical models in one- (1D) and two-dimensions (2D), we discuss which kinds of impurities do not alter the global topological order represented by spatially averaged topological marker. Remarkably, we discover that weak and dilute impurities that correspond to varying nonzero matrix elements of the lattice Hamiltonian do not alter the average topological marker, regardless whether the impurities represent local variation of hopping, chemical potential, or superconducting pairing, as can be proved analytically in 1D and 2D by a perturbation theory. Such a feature seems to bear a striking similarity to Anderson's theorem for $s$-wave superconductivity\cite{Anderson59}. On the other hand, if the impurities correspond to varying zero matrix elements of the lattice Hamiltonian, then the average topological marker is generally not conserved. In addition, for strong and dense impurities, the perturbation theory fails, and we find that if the impurity strength exceeds the band gap and the impurity density exceeds that set by a correlation length, the topological order is destroyed. Moreover, impurities may be used to continuously interpolate the average topological marker from one integer to another, mimicking a first-order phase transition. Various prototype theoretical models of TIs and TSCs, together with a variety of impurities, are employed to demonstrate the ubiquity of our statements.

The structure of the paper is organized in the following manner. In Sec.~\ref{sec:local_deformation_Hamiltonian}, we first briefly review the derivation of topological marker, and elaborate how impurities can be viewed as local deformation of a lattice Hamiltonian. In Sec.~\ref{sec:1D_topo_materials}, we investigate 1D class BDI and D systems, provide an analytical proof for the conservation of average topological marker, and numerically demonstrate the continuous interpolation between different topological phases caused by impurities. In Sec.~\ref{sec:2D_topo_materials}, we focus on 2D class A, AII and D to elaborate the same features in 2D systems. Finally, the results will be summarized in Sec.~\ref{sec:conclusions}.




\section{Local deformation of lattice Hamiltonian \label{sec:local_deformation_Hamiltonian}}

\subsection{Proper choice the position operator and the average topological marker}

We start by giving a brief overview of the universal topological marker within the context of symmetry classification. The systems under consideration are the TIs and TSCs in $D$-dimension described by Dirac Hamiltonian $H={\bf d}({\bf k})\cdot{\boldsymbol\Gamma}$, where ${\bf d}({\bf k})=(d_{0},d_{1}...d_{D})$ characterizes the momentum dependence of the Hamiltonian, and $\Gamma_{i}=(\Gamma_{0},\Gamma_{1}...\Gamma_{2n})$ are the $n$-th order Dirac matrices of dimension $2^{n}\times 2^{n}$ that satisfy $\left\{\Gamma_{i},\Gamma_{j}\right\}=2\delta_{ij}$\cite{Schnyder08,Ryu10,Chiu16}. The Hamiltonian is classified according to the TR, particle-hole (PH) and chiral symmetries, yet the topological invariant in any dimension and symmetry class can be constructed in a unified manner from the unit vector ${\bf n(k)=d(k)/|d(k)|}$ of the Dirac Hamiltonian\cite{vonGersdorff21_unification}, which has been called the wrapping number or degree of the map ${\rm deg}[{\bf n}]$. The universal topological marker is subsequently derived from rewriting the momentum space topological invariant into real space according to the following recipe\cite{Chen23_universal_marker}. Suppose for a specific TI or TSC, the Dirac Hamiltonian uses only $\left\{\Gamma_{0},\Gamma_{1},...\Gamma_{D}\right\}$, leaving $\left\{\Gamma_{D+1},\Gamma_{D+2},...\Gamma_{2n}\right\}$ unused. We denote the product of all the unused ones to be $W=\Gamma_{D+1}\Gamma_{D+2}...\Gamma_{2n}$, or $W=I$ if all Dirac matrices are used, and introduce the projectors to the filled and empty lattice eigenstates
\begin{eqnarray}
&&\sum_{m}|m\rangle\langle m|\equiv Q,\;\;\;\sum_{n}|n\rangle\langle n|\equiv P,
\label{projector_lattice_eigenstates}
\end{eqnarray}
where $|n\rangle$ is the filled eigenstate of negative energy $E_{n}<0$, and $|m\rangle$ is the empty eigenstate of positive energy $E_{m}>0$ obtained from diagonalizing the lattice Hamiltonian $H_{0}|\ell\rangle=E_{\ell}|\ell\rangle$. The projector formalism leads to a real space expression for the momentum space topological invariant provided the system remains homogeneous\cite{Chen23_universal_marker}
\begin{eqnarray}
{\rm Homogeneous\;\;systems}:\;\;\;{\rm deg}[{\bf n}]=\frac{1}{L^{D}}{\rm Tr}[\,{\hat{\cal C}}\,].
\label{degn_TrC_correspondence}
\end{eqnarray}
where $L^{D}$ is the total number of unit cells, and ${\rm Tr}[...]$ represents the trace over all the internal degrees of freedom on all lattice sites. The topological operator ${\hat{\cal C}}$ in Eq.~(\ref{degn_TrC_correspondence}) consists of alternating projectors and position operators 
\begin{eqnarray}
{\hat {\cal C}}=N_{D}W\left[Q\,{\hat i_{1}}P\,{\hat i_{2}}...\,{\hat i_{D}}{\cal O}+(-1)^{D+1}P\,{\hat i_{1}}Q\,{\hat i_{2}}...{\hat i_{D}}{\overline{\cal O}}\right],\;\;\;\;\;
\label{topological_operator}
\end{eqnarray}
where $\left\{{\hat i}_{1},{\hat i}_{2},{\hat i}_{3}...\right\}=\left\{{\hat x},{\hat y},{\hat z}...\right\}$, and the last operators are $\left\{{\cal O},\overline{\cal O}\right\}=\left\{P,Q\right\}$ if $D=$ odd, and $\left\{{\cal O},\overline{\cal O}\right\}=\left\{Q,P\right\}$ if $D=$ even owing to the alternating ordering of the projectors $Q$ and $P$. The normalization factor is given by $N_{D}=i^{D}2^{2D-n}\pi^{D}/c\,V_{D}$, where $V_{D}=2\pi^{(D+1)/2}/\Gamma(\frac{D+1}{2})$ is the volume of the $D$-sphere of unit radius that takes the value $\left\{V_{1},V_{2},V_{3}...\right\}=\left\{2\pi,4\pi,2\pi^{2}...\right\}$, and the factor $c=\left\{1,-1,i,-i\right\}$ is defined from the trace of all the $\Gamma$-matrices multiplied together ${\rm Tr}\left[\Gamma_{0}\Gamma_{1}...\Gamma_{2n}\right]=2^{n}c$, which depends on the representation of the $\Gamma$-matrices for the system at hand. The topological marker on a lattice site ${\bf r}$ is then defined as the ${\bf r}$-th diagonal element of topological operator
\begin{eqnarray}
{\cal C}({\bf r})=\langle{\bf r}|{\hat{\cal C}}|{\bf r}\rangle=\sum_{\sigma}\langle{\bf r}\sigma|{\hat{\cal C}}|{\bf r}\sigma\rangle,
\end{eqnarray}
where $\sum_{\sigma}$ represents the summation over all the $2^{n}$ internal degrees of freedom inside the unit cell at ${\bf r}$, such as spin, orbital, particle-hole, etc. 


We now comment on the choice of position operators $\left\{{\hat i}_{1},{\hat i}_{2}...{\hat i}_{D}\right\}$ in Eq.~(\ref{topological_operator}). In a previous work\cite{Chen23_universal_marker}, we employed the straight forward way of assigning the position operators by simply numerating the position of the unit cell. For instance, in a 1D system of $L$ unit cells, we have used 
\begin{eqnarray}
{\hat i}_{1}={\hat x}={\rm diag}\left(1,2,3...L\right)\otimes I_{\sigma},
\label{position_operator_1234}
\end{eqnarray}
where $I_{\sigma}$ is a $\sigma\times\sigma$ identity matrix, since all the internal degrees of freedom $\sigma$ within the same unit cell should be assigned by the same position operator. However, this position operator does not respect the periodic boundary condition (PBC), since there is a sharp jump from $L$ back to the first site $1$, leading to an anomaly on the spatial profile of ${\cal C}({\bf r})$ at the two ends $r=1$ and $r=L$. In other words, the topological marker is not quantized to integer everywhere on a system with PBC, a known numerical artifact\cite{Bianco11,Chen23_universal_marker} that makes it difficult to examine the variation of topological marker caused by impurities.

To fix this problem, in the present work we assign the position operator by
\begin{eqnarray}
{\hat i}_{1}={\hat x}=\frac{L}{2\pi}{\rm diag}\left(e^{2\pi i/L},e^{4\pi i/L},e^{6\pi i/L}...e^{2\pi i}\right)\otimes I_{\sigma},\;\;\;
\label{position_operator_exp}
\end{eqnarray}
and similarly for all other spatial directions, i.e., using the enumeration $\frac{L}{2\pi}\exp(2\pi ix/L)$ in the position operator, a form that respects PBC as has been proposed previously\cite{Aligia99,Prodan10,Aligia23,Molignini23_TPT_finite_T}. The trade off is that the marker will become complex since the position operator is complex. Thus we choose to use Eq.~(\ref{position_operator_exp}) and examine only the modulus of the topological marker 
\begin{eqnarray}
|{\cal C}({\bf r})|=\left[({\rm Re}{\cal C}({\bf r}))^2+({\rm Im}{\cal C}({\bf r}))^2\right]^{1/2},
\label{modulus_Cr}
\end{eqnarray}
which is found to be quantized to integer everywhere in the system and hence suitable to answer whether impurities modify the average topological marker. We further defined the average topological marker by
\begin{eqnarray}
{\overline{\cal C}}\equiv\frac{1}{L^{D}}\sum_{\bf r}|{\cal C}({\bf r})|.
\end{eqnarray}
In a homogeneous system, the topological marker on every site is equal to the average topological marker, which is also equal to the modulus of the momentum space topological invariant $|{\cal C}({\bf r})|=\overline{\cal C}=|{\rm deg}[{\bf n}]|$. The key issue we want to address in the present work is the condition under which the average topological marker $\overline{\cal C}$ remains unchanged from $|{\rm deg}[{\bf n}]|$ even in the presence of impurities, despite the local marker $|{\cal C}({\bf r})|$ can vary in space. For simplicity of the notation, in what follows we use ${\cal C}({\bf r})$ and $|{\cal C}({\bf r})|$ interchangeably, although one should keep in mind that we are actually presenting the modulus of the topological marker $|{\cal C}({\bf r})|$.



\subsection{Disorder as local deformation of lattice Hamiltonian }

Real materials contain all sorts of defects. Even if no lattice sites are missing, the local energetic parameters like chemical potential, hopping, or pairing, can still vary because of impurities. Remarkably, we find that despite different kinds of impurities represent different physical quantities, whether the average marker ${\overline{\cal C}}$ remains conserved in the presence of impurities only depends on whether they correspond to local variation of zero or nonzero matrix elements of the lattice Hamiltonian $H_{0}$. As we shall elaborate in the following sections, particularly for the disorder that corresponds to changing a nonzero matrix element $(H_{0})_{ij}=(H_{0})_{ji}^{\ast}\equiv\lambda$ to $\lambda+\delta\lambda$, analytical formula can be derived to demonstrate the invariance of average topological marker. The full Hamiltonian with such a perturbation is
\begin{eqnarray}
H=H_{0}+\delta\lambda\,\partial_{\lambda}H_{0}.
\label{H0_dlambdaH0}
\end{eqnarray}
The $\partial_{\lambda}H_{0}$ is a matrix where the perturbed element and its conjugate are $1$ and all other matrix elements are zero. Furthermore, one can easily generalize the following argument of perturbation theory to simultaneously varying multiple nonzero matrix elements $\left\{\lambda_{1},\lambda_{2}...\right\}$ such that $H=H_{0}+\sum_{\alpha}\delta\lambda_{\alpha}\,\partial_{\lambda_{\alpha}}H_{0}$, which correspond to the case of an extended impurity or multiple impurities. 

Treating the variation $\delta\lambda\partial_{\lambda}H_{0}$ in Eq.~(\ref{H0_dlambdaH0}) as a perturbation and expand the eigenstates to leading order yields 
\begin{eqnarray}
|\ell '\rangle=|\ell\rangle+\delta\lambda\sum_{k\neq\ell}\frac{|k\rangle\langle k|\partial_{\lambda}H_{0}|\ell\rangle}{E_{\ell}-E_{k}},
\label{eigenstate_perturbation}
\end{eqnarray}
valid for either the filled $|\ell\rangle=|n\rangle$ or the empty $|\ell\rangle=|m\rangle$ states. As a result, the projectors to the filled and empty states in Eq.~(\ref{projector_lattice_eigenstates}) are modified by $P\rightarrow P'=\sum_{n'}|n'\rangle\langle n'|$ and $Q\rightarrow Q'=\sum_{m'}|m'\rangle\langle m'|$, respectively. Expanding $P'$ to leading order in $\delta\lambda$ yields
\begin{eqnarray}
&&P'=P+\delta\lambda\sum_{n}\sum_{k\neq n}\left(\frac{|n\rangle\langle n|\partial_{\lambda}H_{0}|k\rangle\langle k|}{E_{n}-E_{k}}+h.c.\right)
\nonumber \\
&&=P+\delta\lambda\sum_{n}\sum_{k\neq n}\left(|n\rangle\langle\partial_{\lambda}n|k\rangle\langle k|+|k\rangle\langle k|\partial_{\lambda}n\rangle\langle n|\right)
\nonumber \\
&&=P+\delta\lambda\sum_{n}\left(|n\rangle\langle\partial_{\lambda}n|+|\partial_{\lambda}n\rangle\langle n|\right)
\nonumber \\
&&=P+\delta\lambda\partial_{\lambda}P,
\label{Pprime_P_dP}
\end{eqnarray}
where we have used $\sum_{k\neq n}|k\rangle\langle k|=I-|n\rangle\langle n|$. The same argument also leads to $Q'=Q+\delta\lambda\partial_{\lambda}Q$, meaning that the corrections to the projectors are simply given by taking the derivative $\partial_{\lambda}$ on the projectors. As a result, the topological invariant itself can be expanded by
\begin{eqnarray}
{\cal C}'={\cal C}+\delta\lambda\partial_{\lambda}{\cal C},
\label{Cp_C_dlambdaC}
\end{eqnarray}
where $\partial_{\lambda}{\cal C}$ contains straightforward derivatives of operators $P$ and $Q$. If $\partial_{\lambda}{\cal C}=0$ then the average topological marker remains unchanged ${\cal C}'={\cal C}$ in the presence of the disorder $\delta\lambda$, which is what we aim to prove. We shall see below how this occurs in each symmetry class in 1D and 2D.

\section{One-dimensional topological materials \label{sec:1D_topo_materials}}

The topological operator in 1D is given by
\begin{eqnarray}
{\hat{\cal C}}_{1D}=N_{D}W\left[Q{\hat x}P+P{\hat x}Q\right].
\label{1D_topological_operator}
\end{eqnarray}
For the symmetry classes AIII, BDI, CII, DIII in 1D that preserve chiral symmetry, the $W$ matrix is proportional to the chiral operator $W\propto S$, whereas the class D has a different interpretation of $W$. It turns out that the average topological marker ${\cal C}$ remains unchanged in all 1D systems because of the same reason, namely the swapping of projectors $P$ and $Q$ under the multiplication of $W$, as we shall explain below.

\subsection{1D class BDI \label{sec:1D_class_BDI}}

For concreteness, we use the prototype spinless Su-Schrieffer-Heeger (SSH) model as an example for 1D class BDI, described by the lattice Hamiltonian~\cite{Su79} 
\begin{eqnarray}
{\cal H}_{0}&=&\sum_{i}(t+\delta t)c_{Ai}^{\dag}c_{Bi}+(t-\delta t)c_{Ai+1}^{\dag}c_{Bi}+h.c.
\label{SSH_lattice_model}
\end{eqnarray}
where $i$ denotes the unit cell position with two sublattices $A$ and $B$ with the corresponding fermion annihilation operators $c_{Ai}$ and $c_{Bi}$, respectively, and $t\pm\delta t$ are the hopping amplitudes that alternate between even and odd bonds. The lattice Hamiltonian satisfies the chiral symmetry 
\begin{eqnarray}
SH_{0}S^{-1}=SH_{0}S=-H_{0},
\end{eqnarray}
where the chiral operator $S=I_{N\times N}\otimes\sigma_{z}$ is constructed by enlarging $\sigma_{z}$ to the lattice degrees of freedom of $N$ unit cells of $2$ sublattices, which is also the $W=W^{-1}=S$ matrix in the topological operator, and moreover the prefactor in Eq.~(\ref{1D_topological_operator}) is unity $N_{D}=1$. The projectors in Eq.~(\ref{projector_lattice_eigenstates}) transform under chiral operator like 
\begin{eqnarray}
WPW^{-1}=Q,\;\;\;WQW^{-1}=P.
\label{WPW_WQW}
\end{eqnarray}
As a result, the traces of the two terms in Eq.~(\ref{1D_topological_operator}) are actually equal
\begin{eqnarray}
&&{\rm Tr}\left[WP{\hat x}Q\right]={\rm Tr}\left[WPW^{-1}W{\hat x}QW^{-1}W\right]
\nonumber \\
&&={\rm Tr}\left[WQ{\hat x}P\right],
\end{eqnarray}
so we may use ${\cal C}=(2N_{D}/L){\rm Tr}\left[WQ{\hat x}P\right]$ to calculate the average topological marker. We find that the derivative of ${\cal C}$ vanishes identically 
\begin{eqnarray}
&&\partial_{\lambda}{\cal C}=\frac{2N_{D}}{L}{\rm Tr}\left[W\partial_{\lambda}Q{\hat x}P+WQ{\hat x}\partial_{\lambda}P\right]
\nonumber \\
&&=\frac{2N_{D}}{L}{\rm Tr}\left[WW^{-1}PW\partial_{\lambda}Q{\hat x}+\partial_{\lambda}PWQW^{-1}W{\hat x}\right]
\nonumber \\
&&=\frac{2N_{D}}{L}{\rm Tr}\left[W(Q\partial_{\lambda}Q+\partial_{\lambda}PP){\hat x}\right]
\nonumber \\
&&=\frac{2N_{D}}{L}{\rm Tr}\left[W(\partial_{\lambda}QP+\partial_{\lambda}PP){\hat x}\right]
\nonumber \\
&&=\frac{2N_{D}}{L}{\rm Tr}\left[W\partial_{\lambda}IP{\hat x}\right]=0,
\label{dC_1D}
\end{eqnarray}
where in the fourth line we have used $\partial_{\lambda}Q=-\partial_{\lambda}P$, and the fact that $QP=0$ leads to $Q\partial_{\lambda}P=-\partial_{\lambda}QP$. As a result, the average topological marker ${\cal C}$ remains unchanged under the deformation $\lambda\rightarrow\lambda+\delta\lambda$ according to Eq.~(\ref{Cp_C_dlambdaC}), where the deformation in this case means locally changing the hopping $\lambda=t\pm\delta t$ on some bonds to a different value $\lambda+\delta\lambda$.

\begin{figure}[ht]
\begin{center}
\includegraphics[clip=true,width=0.99\columnwidth]{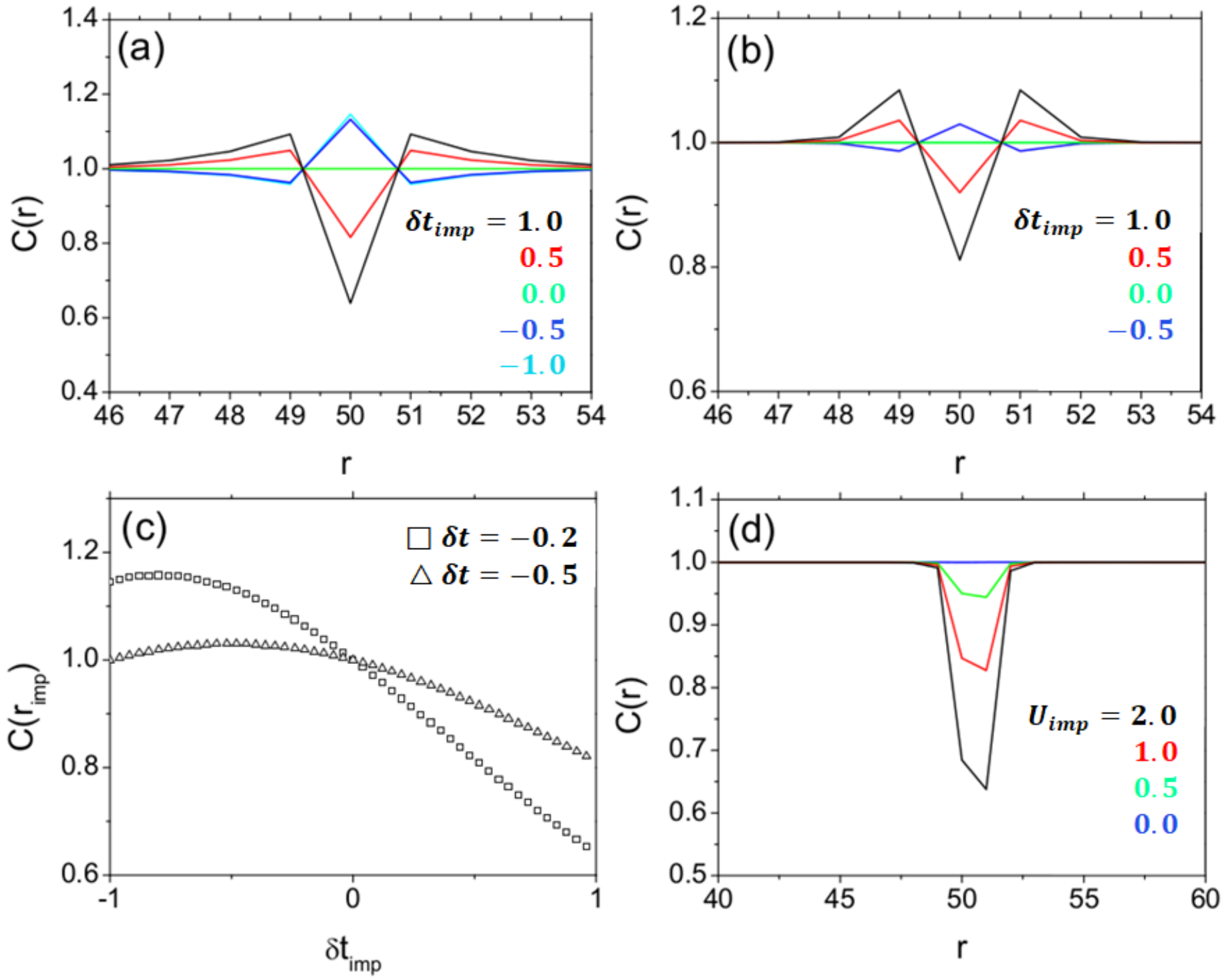}
\caption{Spatial profile of the topological marker ${\cal C}({\bf r})$ at tuning parameter (a) $\delta t=-0.2$ (close to the critical point) and (b) $\delta t=-0.5$ (far from the critical point) near a single hopping impurity $\delta t_{imp}$. For any values of $\delta t_{imp}$, the average topological marker $\overline{\cal C}\approx 1$ remains unchanged. (c) The evolution of the marker exactly on the impurity site as a function of $\delta t_{imp}$. (d) At $\delta t=-0.5$, we examine a potential impurity $U_{imp}$ of different magnitude, which yields an average marker $\overline{\cal C}$ that is not conserved.  } 
\label{fig:SSH_single_imp_data}
\end{center}
\end{figure}

We further use numerical calculation to confirm these statements. In Fig.~\ref{fig:SSH_single_imp_data} (a), we show the spatial profile of the marker ${\cal C}({\bf r})$ around a hopping impurity $\delta t_{imp}$ that changes the hopping on a specific bond, which alters a nonzero element in the Hamiltonian. The marker has a spatial profile that is reduced on the impurity site but enhanced on neighboring sites, rendering the average marker ${\overline{\cal C}}$ conserved, a universal feature for all those cases where ${\overline{\cal C}}$ is conserved. Thus the hopping impurity does alter the spatial profile of ${\cal C}({\bf r})$ around its neighborhood, but for any strength of $\delta t_{imp}$, the average marker $\overline{\cal C}\approx 1$ always remains quantized up to numerical precision, consistent with Eq.~(\ref{dC_1D}). Moreover, by comparing the results at tuning parameter $\delta t=-0.2$ and $\delta t=-0.5$ as shown in Fig.~\ref{fig:SSH_single_imp_data} (a) and (b), the variation around the impurity seems to become more short ranged at the larger tuning parameter $\delta t=-0.5$, suggesting that the correlation length $\xi\propto 1/|\delta t|$, equivalently the decay length of the topological edge state, also characterizes the spatial profile of the marker around an impurity\cite{Chen17,Chen19_AMS_review,Chen19_universality}. Finally, the marker exactly at the impurity site ${\cal C}({\bf r}_{imp})$ shows a strong dependence on the impurity strength $\delta t_{imp}$ as shown in Fig.~\ref{fig:SSH_single_imp_data} (c), but the average marker is still conserved. 


On the other hand, if one considers a point-like potential impurity $U_{imp}$ that changes the chemical potential on a specific site, then it corresponds to locally varying a zero matrix element since the SSH model in Eq.~(\ref{SSH_lattice_model}) does not have a chemical potential to begin with. Note that this kind of potential impurity is actually a practical issue, since doping is what makes polyacetylene a conductive polymer and useful for commercial purposes\cite{Heeger88}. In this case, the variation of ${\cal C}({\bf r})$ around the impurity shown in Fig.~\ref{fig:SSH_single_imp_data} (d) reduces the average marker $\overline{\cal C}$, making it not conserved. We have also tried other kinds of impurities that correspond to varying zero matrix elements, such as a local neat-nearest-neighbor hopping (not shown), and also obtained nonconserved average marker $\overline{\cal C}$.

\begin{figure}[ht]
\begin{center}
\includegraphics[clip=true,width=0.9\columnwidth]{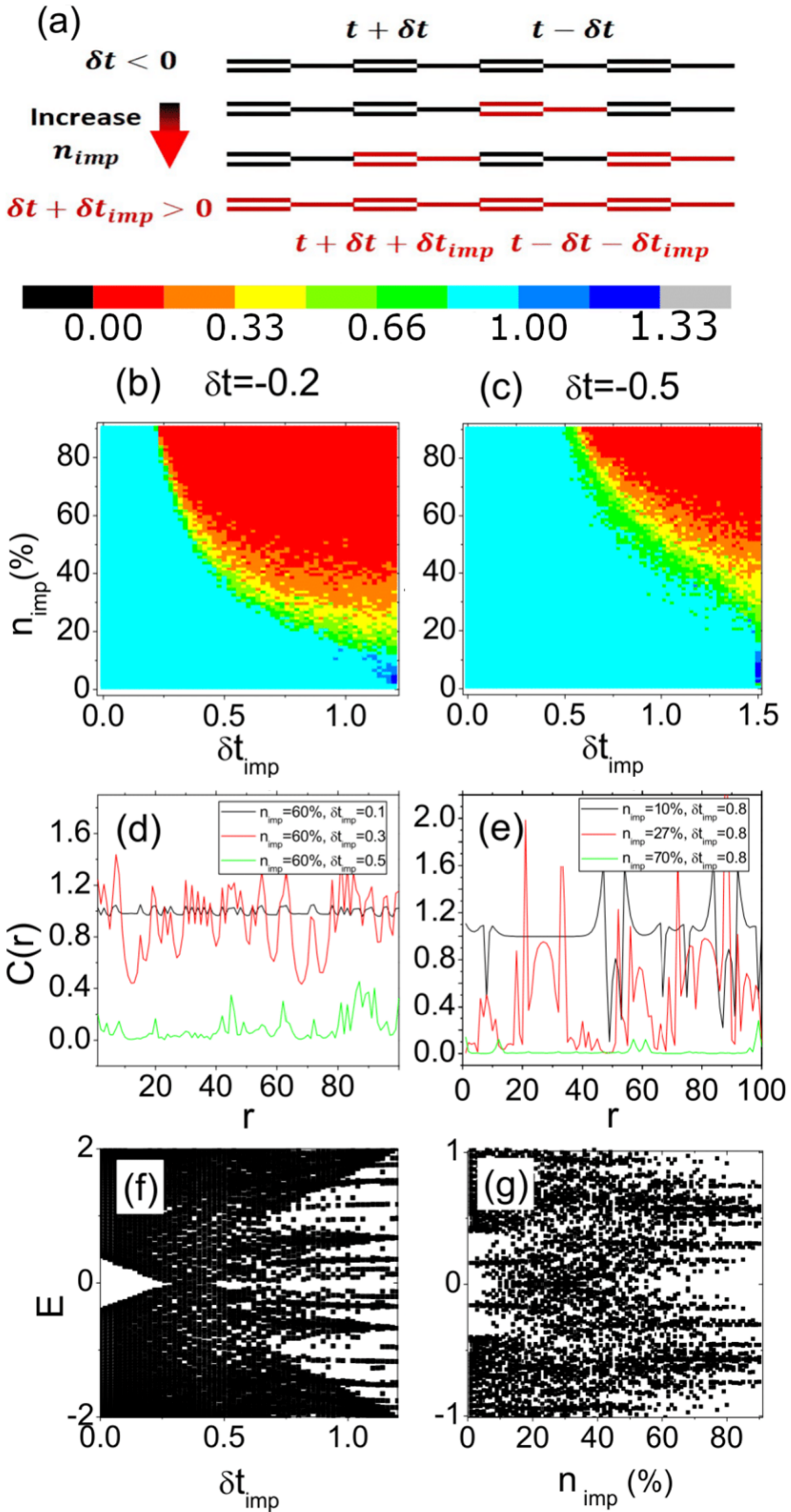}
\caption{(a) Schematics of our disordered SSH model with hopping impurities that continuously interpolates between the trivial and nontrivial phases. The average marker $\overline{\cal C}$ plotted as a function of the strength $\delta t_{imp}$ and density $n_{imp}$ of the hopping impurities is shown for (b) the tuning parameter $\delta t=-0.2$ close to the critical point, and (c) $\delta t=-0.5$ far from the critical point. Panels (d) and (e) show the spatial profile of the marker ${\cal C}({\bf r})$ as the system crosses into the trivial phase upon increasing impurity strength $\delta t_{imp}$ and density $n_{imp}$, respectively. Panels (f) and (g) show the evolution of eigenenergies across the transition by increasing $\delta t_{imp}$ and $n_{imp}$, respectively, where one sees that the crossover region has a gapless energy spectrum.  } 
\label{fig:SSH_many_imp_result}
\end{center}
\end{figure}

On the other hand, our analytical proof given in Eq.~(\ref{dC_1D}) is expected to fail if there are many impurities and the impurity density becomes too high, since the first-order perturbation theory in Eq.~(\ref{eigenstate_perturbation}) is no longer adequate. To examine the situation of many impurities, we rely on numerical calculation to determine whether the average marker $\overline{\cal C}$ remains conserved. In particular, we are interested in the issue of whether many hopping impurities can bring the system continuously from a nontrivial to a trivial phase without going through a sharp transition, thereby reminiscing a first-order topological phase transition (TPT). For this purpose, we examine the following disordered SSH model
\begin{eqnarray}
{\cal H}&=&\sum_{i}(t+\delta t)c_{Ai}^{\dag}c_{Bi}+(t-\delta t)c_{Ai+1}^{\dag}c_{Bi}
\nonumber \\
&&+\sum_{i\in imp}\delta t_{imp}\left(c_{Ai}^{\dag}c_{Bi}-c_{Ai+1}^{\dag}c_{Bi}\right)+h.c.
\label{disordered_SSH_model}
\end{eqnarray}
where $i\in imp$ represents the impurity unit cells. The model is shown schematically in Fig.~\ref{fig:SSH_many_imp_result} (a). In the absence of the impurities $\delta t_{imp}=0$, a negative tuning parameter $\delta t$ makes the unperturbed SSH model topologically nontrivial $\overline{\cal C}=1$. Naively, if we have many positive hopping impurities $\delta t_{imp}>0$ and moreover $\delta t_{imp}$ is so large such that it overcomes the negative tuning parameter, then one expects that the system should become topologically trivial. This is consistent with the high impurity density $n_{imp}\approx 100\%$ and strong positive hopping $\delta t_{imp}>1.2$ limit of the phase diagram of Fig.~\ref{fig:SSH_many_imp_result} (b), which has $\overline{\cal C}=0$. The interesting issue is then how the average marker $\overline{\cal C}$ behaves as one continuously changes $\delta t_{imp}$ from negative to positive and increases the impurity density $n_{imp}$. Remarkably, we find that there is no sharp jump for $\overline{\cal C}$ as a function of $\left\{\delta t_{imp},n_{imp}\right\}$, but rather a smooth crossover between the nontrivial $\overline{\cal C}=1$ and the trivial $\overline{\cal C}=0$ phases, thereby exhibiting the feature of a first-order TPT. Comparing the phase diagrams at the tuning parameter $\delta t=-0.2$ (close to the critical point) in Fig.~\ref{fig:SSH_many_imp_result} (b) and at $\delta t=-0.5$ (far from the critical point) in Fig.~\ref{fig:SSH_many_imp_result} (c), we see that the later requires much stronger magnitude $\delta t_{imp}$ and density $n_{imp}$ to reach the trivial phase $\overline{\cal C}=0$. This is very intuitive, since $\delta t=-0.5$ is deep inside the topological phase and hence requires a larger perturbation to break the topological order, whereas $\delta t=-0.2$ is closer to the critical point and hence a small perturbation can already drive it across the phase boundary. Empirically, we find that the topological order is destroyed if the average distance between impurities is shorter than the correlation length set by Fermi velocity divided by the band gap $\xi =v_{F}/|M|$, and if the impurity strength $\delta\lambda$ is higher than the band gap
\begin{eqnarray}
\frac{L}{n_{imp}N}\apprle\xi,\;\;\;\delta\lambda\apprge |M|,
\label{emperical_many_impurity_formula}
\end{eqnarray}
where $N$ is the number of unit cells, $L$ is the length of the system. In this disordered SSH model example we have $M=2\delta t$, $\delta\lambda=\delta t_{imp}$, and $\xi\sim at/|\delta t|$.

To further understand the nature of this smooth crossover from the nontrivial to trivial phase, in Fig.~\ref{fig:SSH_many_imp_result} (d) we plot the spatial profile of the marker ${\cal C}({\bf r})$ at a fixed impurity density $n_{imp}$ and increasing impurity strength $\delta t_{imp}$, and in Fig.~\ref{fig:SSH_many_imp_result} (e) we plot ${\cal C}({\bf r})$ at a fixed $\delta t_{imp}$ and increasing $n_{imp}$. Both figures show that the overall magnitude of the marker gradually reduces, signifying the crossover to the trivial phase $\overline{\cal C}=0$. This result also indicates that it is  possible to fine tune the values of $\delta t_{imp}$ and $n_{imp}$ such that the average marker takes some fractional value $0<\overline{\cal C}<1$. Moreover, since both the trivial $\overline{\cal C}=0$ and the nontrivial $\overline{\cal C}=1$ phases are usually characterized by a gapped energy spectrum, it is intriguing to ask what happens to the energy spectrum in this crossover regime $0<\overline{\cal C}<1$. Interestingly, as shown in Fig.~\ref{fig:SSH_many_imp_result} (f) and (g), we find that indeed both the $\overline{\cal C}=0$ and $\overline{\cal C}=1$ phases of our disordered SSH model are gapped, but the crossover region $0<\overline{\cal C}<1$ exhibits a gapless spectrum. This suggests that the disorder can cause the system to smoothly cross between trivial and nontrivial phases by going through a gapless crossover regime, in contrast to the discrete jump of $\overline{\cal C}$ at a second-order TPT in homogeneous models.

In Appendix \ref{apx:crossover_SSH}, we further elaborate how the width of the crossover region in Fig.~\ref{fig:SSH_many_imp_result} (b) depends on $n_{imp}$, as well as showing the evidence that the width remains in the thermodynamic limit $L\rightarrow\infty$. Finally, we remark that because our method only allows to investigate the modulus of the marker $|{\cal C}({\bf r})|$, it will not be able to capture a negative marker caused by impurities in the topologically trivial phase $\overline{\cal C}=0$, shall that happen. This implies that the $\overline{\cal C}=0$ phase in reality may be wider than the red regions on the phase diagrams of Fig.~\ref{fig:SSH_many_imp_result} (b) and (c) (and possibly in all the phase diagrams in the following sections), but our method does not allow to detect this. This issue will require a certain improvement of our formalism in Eq.~(\ref{modulus_Cr}) that currently remains unclear to us, and awaits to be further explored.

\subsection{1D class D}

\begin{figure}[ht]
\begin{center}
\includegraphics[clip=true,width=0.99\columnwidth]{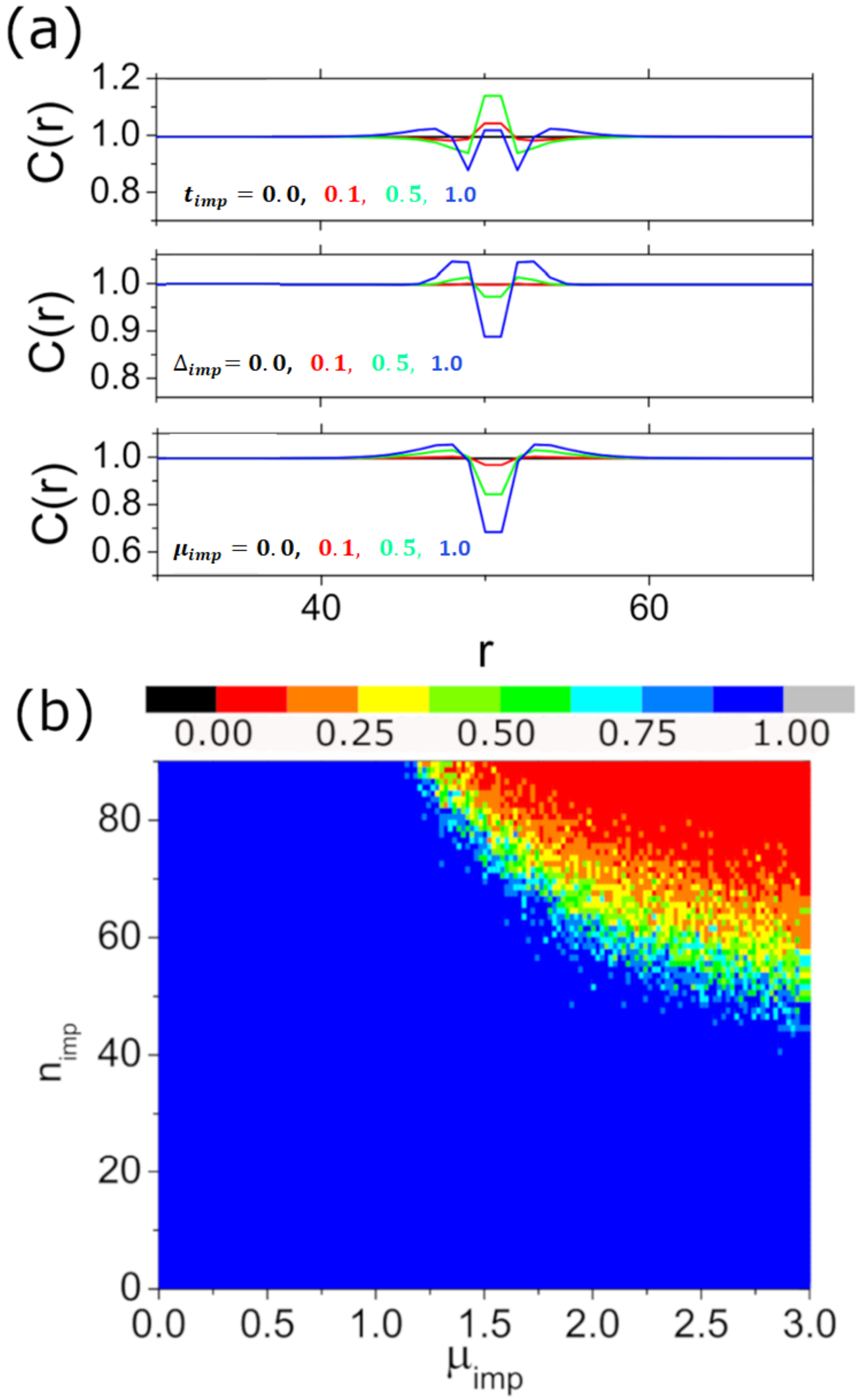}
\caption{(a) Spatial profile of ${\cal C}({\bf r})$ for the Kitaev chain with three different types of impurities from top to bottom: local variation of hopping $t_{imp}$, pairing $\Delta_{imp}$, and chemical potential $\mu_{imp}$. (b) The average marker $\overline{\cal C}$ in the presence of multiple impurities in the chemical potential $\mu_{imp}$ plotted as a function of impurity strength and density.} 
\label{fig:Kitaev_figure}
\end{center}
\end{figure}

A prototype example for 1D class D is the spinless $p$-wave SC chain, or Kitaev chain, described by\cite{Kitaev01}
\begin{eqnarray}
&&H=\sum_{i}t\left(c_{i}^{\dag}c_{i+1}+c_{i+1}^{\dag}c_{i}\right)-\mu \sum_{i}c_{i}^{\dag}c_{i}
\nonumber \\
&&+\sum_{i}\Delta \left(c_{i}c_{i+1}+c_{i+1}^{\dag}c_{i}^{\dag}\right),
\label{Majorana_lattice_model}
\end{eqnarray}
where $c_{i}$ is the spinless fermion annihilation operator at site $i$. The PH symmetry is interpreted by $C=\sigma_{x}K=C^{-1}$, and the Pauli matrix that has not been used is $W=\sigma_{x}=CK$, and the normalization factor is $N_{D}=1$. Note that the chiral symmetry implies that if $|\psi_{\ell k}\rangle$ is an eigenstate of $H(k)$ of eigenenergy $\varepsilon_{\ell}(k)$, then $C|\psi_{\ell k}\rangle$ is an eigenstate of $H(-k)$ of eigenenergy $-\varepsilon_{\ell}(k)$. Combining this with Eq.~(\ref{projector_lattice_eigenstates}) leads to the conclusion that the projectors transform under the PH opertor as
\begin{eqnarray}
&&CPC^{-1}=\int\frac{d^{D}{\bf k}}{(2\pi)^{D}}\sum_{v}C|\psi_{v,k}\rangle\langle \psi_{v,k}|C^{-1}
\nonumber \\
&&=\int\frac{d^{D}{\bf k}}{(2\pi)^{D}}\sum_{c}|\psi_{c,-k}\rangle\langle \psi_{c,-k}|=Q,
\end{eqnarray}
and likewisely $CQC^{-1}=P$. Together with the fact that $P=P^{\dag}$ and $Q=Q^{\dag}$ are real matrices, we conclude that Eq.~(\ref{WPW_WQW}) still holds in 1D class D with the interpretation $W=\sigma_{x}$. As a result, the proof of $\partial_{\lambda}{\cal C}=0$ in Sec.~\ref{sec:1D_class_BDI} still holds, and hence the average topological marker in 1D class D is unchanged under local variation $\lambda$ that may be either the hopping $t_{imp}$, chemical potential $\mu_{imp}$, or $p$-wave pairing $\Delta_{imp}$ in Eq.~(\ref{Majorana_lattice_model}), as elaborated in Fig.~\ref{fig:Kitaev_figure} (a) with homogeneous parameters fixed at $t=1.0$, $\mu=1.8$, $\Delta=0.5$. Furthermore, Fig.~\ref{fig:Kitaev_figure} (b) also indicates that if these impurities are too many and their strength exceeds the band gap, then the average marker can be destroyed, and the system can undergo a first-order phase transition to the topologically trivial phase just like that shown in Fig.~\ref{fig:SSH_many_imp_result} for the SSH model, suggesting the validity of our empirical formula in Eq.~(\ref{emperical_many_impurity_formula}) for this symmetry class as well.


\section{Two-dimensional topological materials \label{sec:2D_topo_materials}}

The topological operator in 2D has the form
\begin{eqnarray}
{\hat{\cal C}}_{2D}=N_{D}W\left[Q{\hat x}P{\hat y}Q-P{\hat x}Q{\hat y}P\right].
\label{2D_topological_operator}
\end{eqnarray}
For classes A, C, and D that break TR symmetry, the topological marker recovers the well-known the Chern marker\cite{Bianco11} described by $W\propto I$, and for the TR-symmetric classes AII and DIII it recovers the spin Chern marker $W\propto\sigma_{z}$. The invariance of global Chern number in the presence of disorder can be proved analytically, and the invariance of spin Chern number also follows immediately because the Chern number in each spin channel is invariant, as we shall see below for classes A, AII, and D.

\begin{figure}[ht]
\begin{center}
\includegraphics[clip=true,width=0.99\columnwidth]{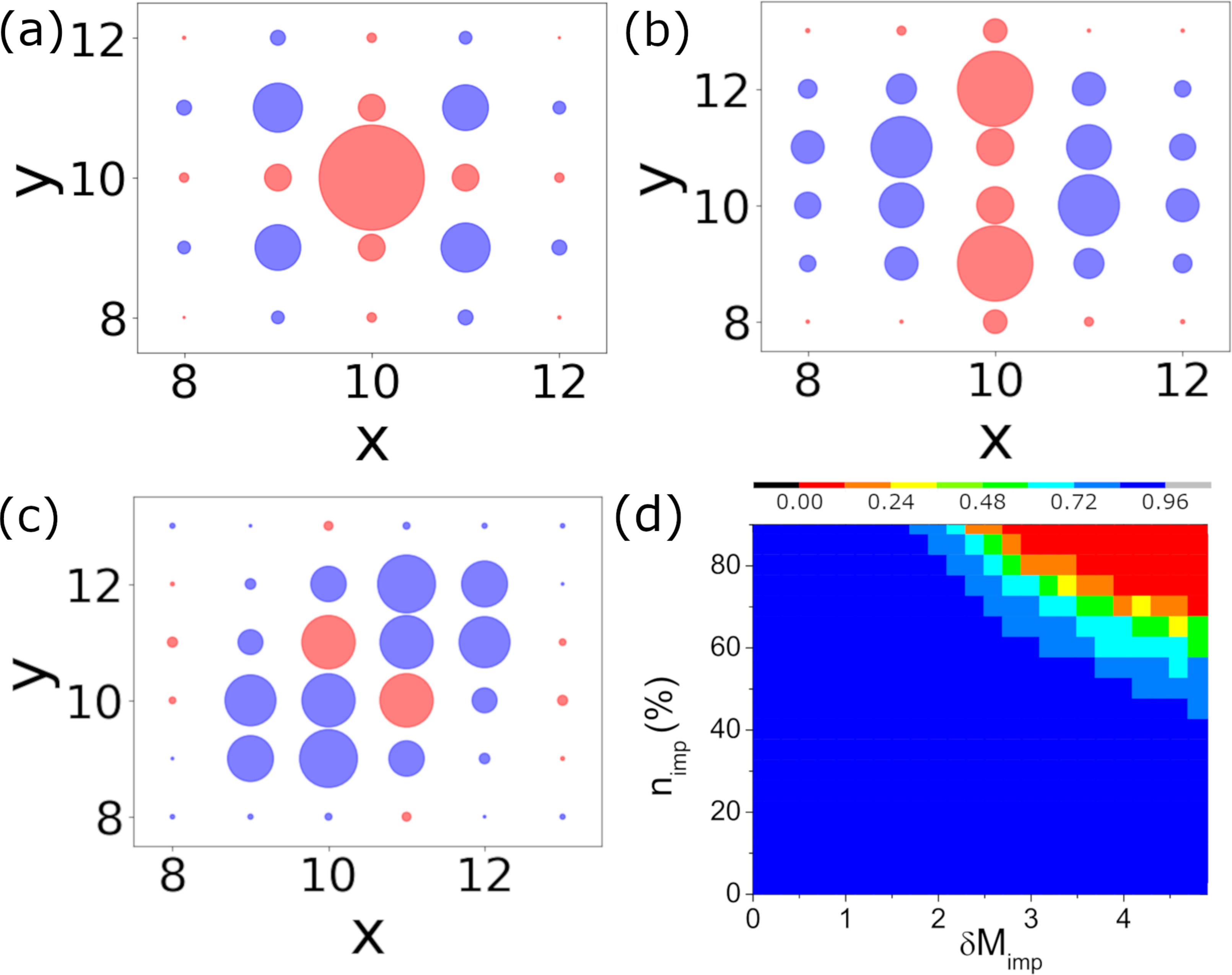}
\caption{The spatial profile of the deviation of local marker ${\cal C}({\bf r})-{\cal C}(\infty)$ from its homogeneous value in the Chern insulator caused by (a) a potential impurity $\mu_{imp}=-5$ ($\overline{\cal C}$ conserved), (b) a hopping impurity $t_{imp}=-0.5$ ($\overline{\cal C}$ conserved), and (c) a next-nearest-neighbor hopping impurity $t_{NNN}=5.0$ ($\overline{\cal C}$ not conserved). The size of the disk represents the magnitude and the color represents the sign (blue$=$positive and red$=$negative, same color code for all the 2D marker figures below), and the axes labels show that coordinates ${\bf r}=(x,y)$ in a $L^{2}=20\times 20$ lattice where the impurity is located at the center. The largest disks in these three figures correspond to ${\cal C}({\bf r})-{\cal C}(\infty)=\left\{-0.641,-0.00413,0.0542\right\}$, respectively. (d) The average marker $\overline{\cal C}$ in the presence of many impurities that locally varying the mass term $\delta M_{imp}$ at different density $n_{imp}$.} 
\label{fig:2D_Chern_figure}
\end{center}
\end{figure}

\subsection{2D class A \label{sec:2D_class_A}}

For 2D class A, there is no unused Dirac matrices so $W=I$. Applying Eqs.~(\ref{Pprime_P_dP}) and (\ref{Cp_C_dlambdaC}) to Eq.~(\ref{2D_topological_operator}), the derivative of the average topological marker is given by
\begin{eqnarray}
\partial_{\lambda}{\cal C}&=&\frac{N_{D}}{L^{2}}{\rm Tr}\left[\partial_{\lambda}Q{\hat x}P{\hat y}Q+Q{\hat x}\partial_{\lambda}P{\hat y}Q+Q{\hat x}P{\hat y}\partial_{\lambda}Q\right.
\nonumber \\
&&\left.-\partial_{\lambda}P{\hat x}Q{\hat y}P-P{\hat x}\partial_{\lambda}Q{\hat y}P-P{\hat x}Q{\hat y}\partial_{\lambda}P\right]
\nonumber \\
&=&\frac{N_{D}}{L^{2}}{\rm Tr}\left[\partial_{\lambda}(QQ){\hat x}P{\hat y}+Q{\hat x}\partial_{\lambda}P{\hat y}\right.
\nonumber \\
&&\left.-\partial_{\lambda}(PP){\hat x}Q{\hat y}-P{\hat x}\partial_{\lambda}Q{\hat y}\right]
\nonumber \\
&=&\frac{N_{D}}{L^{2}}{\rm Tr}\left[-\partial_{\lambda}P{\hat x}{\hat y}+{\hat x}\partial_{\lambda}P{\hat y}\right]=0,
\label{dlambda_Chern_number}
\end{eqnarray}
where we have used $QQ=Q$, $PP=P$, $P+Q=I$, and $\partial_{\lambda}P=-\partial_{\lambda}Q$, and the fact that the position operators commute. Once again this result implies that the average topological marker, i.e., the Chern number, is invariant under change of nonzero matrix elements in the lattice Hamiltonian.

To support this statement, we examine the lattice model of Chern insulators described by\cite{Chen20_absence_edge_current,Molignini23_Chern_marker} 
\begin{eqnarray}
&&H=\sum_{i}t\left\{-ic_{is}^{\dag}c_{i+ap}
+ic_{i+as}^{\dag}c_{ip}+h.c.\right\}
\nonumber \\
&&+\sum_{i}t\left\{-c_{is}^{\dag}c_{i+bp}+c_{i+bs}^{\dag}c_{ip}+h.c.\right\}
\nonumber \\
&&+\sum_{i\delta}t'\left\{-c_{is}^{\dag}c_{i+\delta s}+c_{ip}^{\dag}c_{i+\delta p}+h.c.\right\}
\nonumber \\
&&+\sum_{i}\left(M+4t'\right)\left\{c_{is}^{\dag}c_{is}
-c_{ip}^{\dag}c_{ip}\right\},
\label{Hamiltonian_2DclassA}
\end{eqnarray} 
where $\left\{s,p\right\}$ are the orbitals, $\delta=\left\{a,b\right\}$ denote the lattice constants along planar directions, and $i=\left\{x,y\right\}$ enumerates the planar position.  We considered $t=1.0$, $t'=1.0$ and $M=-2.0$. To highlight the change of marker from the homogeneous value, in Fig.~\ref{fig:2D_Chern_figure} (a) and (b) we present the deviation of the marker ${\cal C}({\bf r})-{\cal C}(\infty)$ from its homogeneous value ${\cal C}(\infty)$ near a potential impurity $\mu_{imp}$ and a hopping impurity $t_{imp}$, respectively, which correspond to varying nonzero elements in the lattice Hamiltonian. Similar to that occurs in 1D, the spatial profile of the marker is such that the reduction on the impurity site ${\cal C}({\bf r})-{\cal C}(\infty)<0$ is compensated by the enhancement on neighboring sites ${\cal C}({\bf r})-{\cal C}(\infty)>0$, rendering the average marker conserved $\sum_{\bf r}\left[{\cal C}({\bf r})-{\cal C}(\infty)\right]\approx 0$ and indicating the validity of Eq.~(\ref{dlambda_Chern_number}). In contrast, in Fig.~\ref{fig:2D_Chern_figure} (c), we put in a local next-nearest-neighbor hopping $t_{NNN}$ that is not in the original Hamiltonian and hence corresponds to varying a zero matrix element, and the resulting average marker is not conserved. Note that these local variations of topological marker correspond to the difference in local heating rate between two circularly polarized lights, which is experimentally measurable in atomic scale by scanning thermal microscopy\cite{Molignini23_Chern_marker}.

We remark that care must be taken when one intends to identify numerically the conservation of the average marker in 2D. Numerically, this can only be done by calculating the difference between the average marker in the presence and in the absence of impurities. For the impurities that are supposed to conserve the marker, this difference is never truly zero in the numerical calculation, but a very small number due to the finite size effect. Thus to firmly identify the conservation of the average marker, a finite size scaling analysis is unavoidable, especially given that the thermodynamic limit is harder to achieve numerically in 2D. For this reason, we have performed such a scaling analysis in Appendix \ref{apx:finite_size_scaling} for all the impurities in all the 2D models investigated in the present work, which confirmed the validity of Eq.~(\ref{dlambda_Chern_number}).

Finally, in Fig.~\ref{fig:2D_Chern_figure} (d), we demonstrate the disorder-induced smooth crossover between the nontrivial $\overline{\cal C}=1$ and trivial $\overline{\cal C}=0$ phases. This crossover is caused by the impurities that correspond to local variation of the mass term $\delta M_{imp}$ that is of opposite sign to the homogeneous value $M$. As a result, at large values of $\delta M_{imp}$ and high impurity density $n_{imp}$, the system must be in the trivial phase by construction. Our numerical result indicates a continuous change from $\overline{\cal C}=1$ to $\overline{\cal C}=0$ as increasing $(\delta M_{imp},n_{imp})$, resembling a first-order transition similar to that discussed in Sec.\ref{sec:1D_class_BDI}. Our results thus suggest that disorder-induced first-order TPT may generally occur in any dimension and symmetry class.

\subsection{2D class AII \label{sec:2D_class_AII}}

A prototype example for 2D class AII that preserves the TR symmetry is the Bernevig-Hughes-Zhang (BHZ) model\cite{Bernevig06,Konig07}, which can be regularized on a square lattice to yield the lattice Hamiltonian\cite{Chen20_absence_edge_current}
\begin{eqnarray}
&&H=\sum_{i\sigma}t\left\{-i\sigma c_{is\sigma}^{\dag}c_{i+ap\sigma}
-i\sigma c_{ip\sigma}^{\dag}c_{i+as\sigma}+h.c.\right\}
\nonumber \\
&&+\sum_{i\sigma}t\left\{-c_{is\sigma}^{\dag}c_{i+bp\sigma}+c_{ip\sigma}^{\dag}c_{i+bs\sigma}+h.c.\right\}
\nonumber \\
&&+\sum_{i\sigma}\left(M+4t'\right)c_{is\sigma}^{\dag}c_{is\sigma}
+\sum_{i}\left(-M-4t'\right)c_{ip\sigma}^{\dag}c_{ip\sigma}
\nonumber \\
&&-\sum_{i\sigma\delta}t'\left\{c_{is\sigma}^{\dag}c_{i+\delta s\sigma}-c_{ip\sigma}^{\dag}c_{i+\delta p\sigma}+h.c.\right\},
\label{Hamiltonian_2DclassAII}
\end{eqnarray}
using the same notation as Eq.~(\ref{Hamiltonian_2DclassA}), and $\sigma=\left\{\uparrow,\downarrow\right\}$ is the spin index. We considered the parameters $t=1.0$, $t'=1.0$, $M=-2.0$. This lattice Hamiltonian is block-diagonal with one block for each spin species, and each block is essentially the lattice Hamiltonian for the Chern insulator in Eq.~(\ref{Hamiltonian_2DclassA}), and so follows the conservation of average spin Chern marker $\overline{\cal C}=(\overline{\cal C}_{\uparrow}-\overline{\cal C}_{\downarrow})/2$, since $\partial_{\lambda}{\cal C}_{\uparrow}=\partial_{\lambda}{\cal C}_{\downarrow}=0$ if the disorder corresponds to varying a nonzero matrix element $\lambda$ according to Eq.~(\ref{dlambda_Chern_number}). It also follows that disorder can be used to induce a first-order TPT in this class, as has been pointed out previously\cite{Shi23}, and moreover the TPT caused by Anderson disorder has been shown recently to exhibit a single-parameter scaling behavior\cite{Assuncao24}.

\begin{figure}[ht]
\begin{center}
\includegraphics[clip=true,width=0.99\columnwidth]{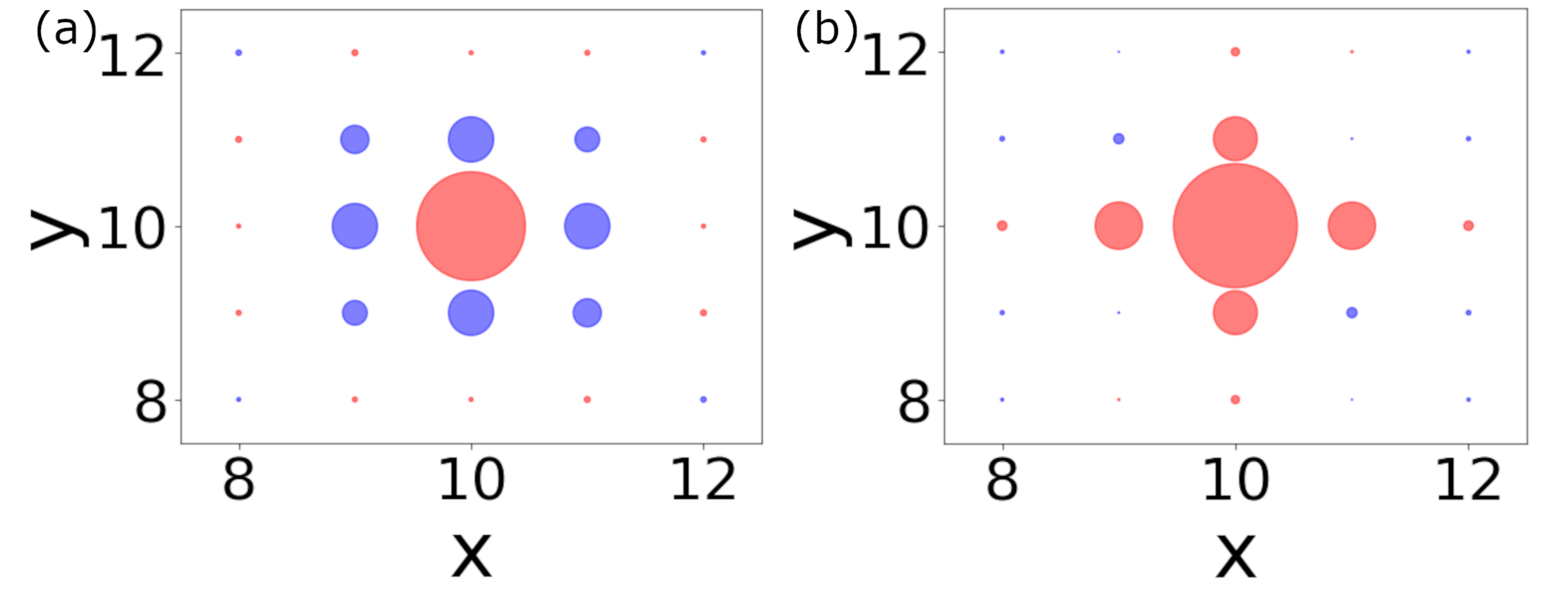}
\caption{The spatial profile of the deviation of local marker from its homogeneous value ${\cal C}({\bf r})-{\cal C}(\infty)$ in the BHZ model presented by size and color of the disks, for a single magnetic impurity of magnitude $S=5.0$ polarized along (a) ${\hat{\bf z}}$ direction (${\overline{\cal C}}$ conserved, largest disk $=-0.53$) and (b) ${\hat{\bf x}}$ direction (${\overline{\cal C}}$ not conserved, largest disk $=-0.64$).  } 
\label{fig:BHZ_figure}
\end{center}
\end{figure}

The investigation of magnetic impurities in the BHZ model helps to demonstrate that breaking the nonspatial symmetry of the system by disorder does not necessarily destroy $\overline{\cal C}$. This statement is made because a magnetic impurity polarized in any direction breaks TR symmetry. However, if the magnetic impurity is polarized along the out-of-plane direction ${\hat{\bf z}}$, then it corresponds to varying nonzero matrix elements (spin $\uparrow$ and $\downarrow$ have opposite corrections to the mass term $M$), and hence $\overline{\cal C}$ should remain conserved according to Eq.~(\ref{dlambda_Chern_number}). In contrast, if the magnetic impurity is polarized in any direction in the $xy$-plane, then it corresponds to varying zero matrix elements and hence $\overline{\cal C}$ may change. This prediction is indeed verified in our numerical calculation of the deviation of the marker shown in Fig.~\ref{fig:BHZ_figure} (a) for the polarization along ${\hat{\bf z}}$ where the deviation sums to zero $\sum_{\bf r}\left[{\cal C}({\bf r})-{\cal C}(\infty)\right]\approx 0$, and Fig.~\ref{fig:BHZ_figure} (b) for the polarization along ${\hat{\bf x}}$ that sums to finite $\sum_{\bf r}\left[{\cal C}({\bf r})-{\cal C}(\infty)\right]\neq 0$. Thus the decisive principle is still whether the disorder varies zero or nonzero matrix elements, while the breaking of nonspatial symmetry may not be detrimental to the conservation of $\overline{\cal C}$.

\subsection{2D class D \label{sec:2D_class_D}}

\begin{figure}[ht]
\begin{center}
\includegraphics[clip=true,width=0.99\columnwidth]{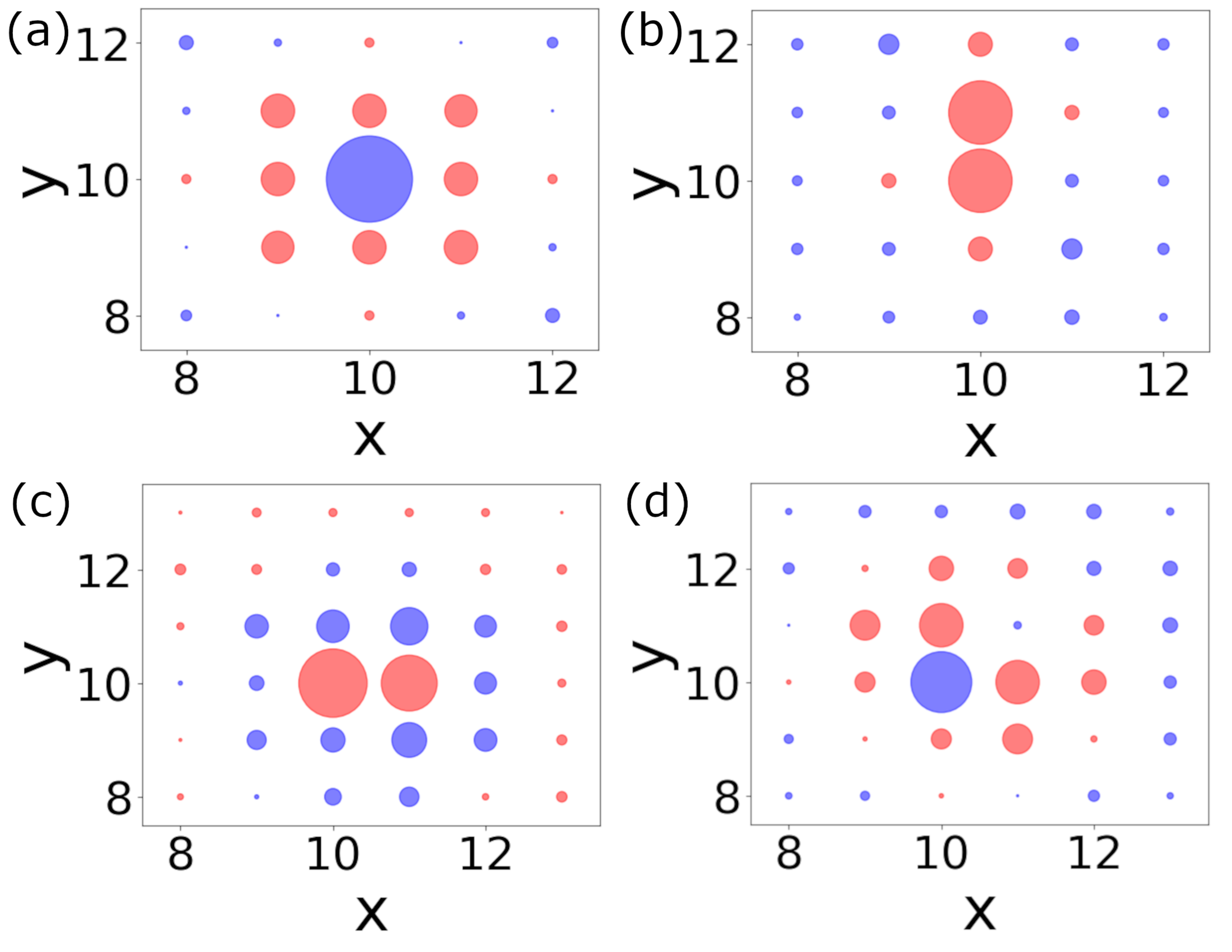}
\caption{The spatial profile of the deviation of the marker from its homogeneous value ${\cal C}({\bf r})-{\cal C}(\infty)$ in chiral $p$-wave TSC caused by a single impurity that correspond to local variation of (a) chemical potential $\mu_{imp}=-1$ (${\overline{\cal C}}$ conserved), (b) pairing $\Delta_{imp}=-0.2$ (${\overline{\cal C}}$ conserved), (c) hopping $t_{imp}=-1.0$ (${\overline{\cal C}}$ conserved), and (d) next-nearest neighbor hopping $t_{NNN}=-1$ (${\overline{\cal C}}$ not conserved). The largest disks in these four figures have values $\left\{0.25,-0.07,-0.1,0.1\right\}$, respectively. } 
\label{fig:pwave_figure}
\end{center}
\end{figure}

For 2D class D, we use the spinless chiral $p$-wave SC as an example\cite{Schnyder08}, described by the lattice Hamiltonian 
\begin{eqnarray}
&&H=\sum_{i\delta}t\left(c_{i}^{\dag}c_{i+\delta}+c_{i+\delta}^{\dag}c_{i}\right)-\mu \sum_{i}c_{i}^{\dag}c_{i}
\nonumber \\
&&+\sum_{i}\Delta \left(-ic_{i}c_{i+x}+ic_{i+x}^{\dag}c_{i}^{\dag}+c_{i}c_{i+y}+c_{i+y}^{\dag}c_{i}^{\dag}\right),
\label{chiral_pwave_lattice_model}
\end{eqnarray}
where $c_{i}$ is the spinless fermion operator at site $i$, $\delta=\left\{x,y\right\}$. We use $\mu=-3.0$, $\Delta=0.5$ and $t=-1.0$. Since all Pauli matrices are used in the $2\times 2$ Hamiltonian, we use $W=I$ and $N_{D}=2\pi i$. The topological marker in this class is precisely the Chern marker in Sec.~\ref{sec:2D_class_A}, so the proof that the Chern number is invariant in the presence of disorder also follows. The numerical results for the deviation of the marker ${\cal C}({\bf r})-{\cal C}(\infty)$ are shown in Fig.~\ref{fig:pwave_figure}, which examine several different kinds of impurities that correspond to varying nonzero matrix elements ($\mu_{imp}$, $\Delta_{imp}$ and $t_{imp}$) indicate a conserved ${\overline{\cal C}}$ in full agreement with Eq.~(\ref{dlambda_Chern_number}). In contrast, the next-nearest-neighbor hopping $t_{NNN}$ that varies a zero matrix element, has a ${\overline{\cal C}}$ that is not conserved.

\subsection{Remarks on three-dimensional topological materials \label{sec:3D_topo_materials}}

Finally, we remark on our current understanding of three-dimensional (3D) TIs and TSCs, which have the following general form of the topological operator
\begin{eqnarray}
{\hat{\cal C}}_{3D}=N_{D}W
\left[Q\hat{x}P\hat{y}Q\hat{z}P+P\hat{x}Q\hat{y}P\hat{z}Q\right].
\end{eqnarray}
The prefactor $N_{D}$ and matrix $W$ for all the 5 nontrivial classes in 3D have all been clarified, and several homogeneous lattice models have already been investigated\cite{Chen23_universal_marker}. However, we are currently unable to provide an analytical proof analogous to Eqs.~(\ref{dC_1D}) and (\ref{dlambda_Chern_number}) on the conditions under which $\overline{\cal C}$ should remain conserved, i.e., what kinds of impurities make the marker conserve and what kinds do not. One may then resort to numerical calculation to seek for the answer, but we find that the exponentiated position operator in Eq.~(\ref{position_operator_exp}) requires a much bigger lattice to make the marker integer (for instance, for a 3D class AII lattice model, the marker reaches $\approx 0.8$ at lattice size $12\times 12\times 12$ at a typical parameter), which makes it difficult to identify the conservation of $\overline{\cal C}$. Thus the investigation of disordered 3D TIs and TSCs, especially whether varying nonzero matrix elements of the Hamiltonian does not alter ${\overline{\cal C}}$, should be left for future investigations.

\begin{figure}[ht]
\begin{center}
\includegraphics[clip=true,width=0.99\columnwidth]{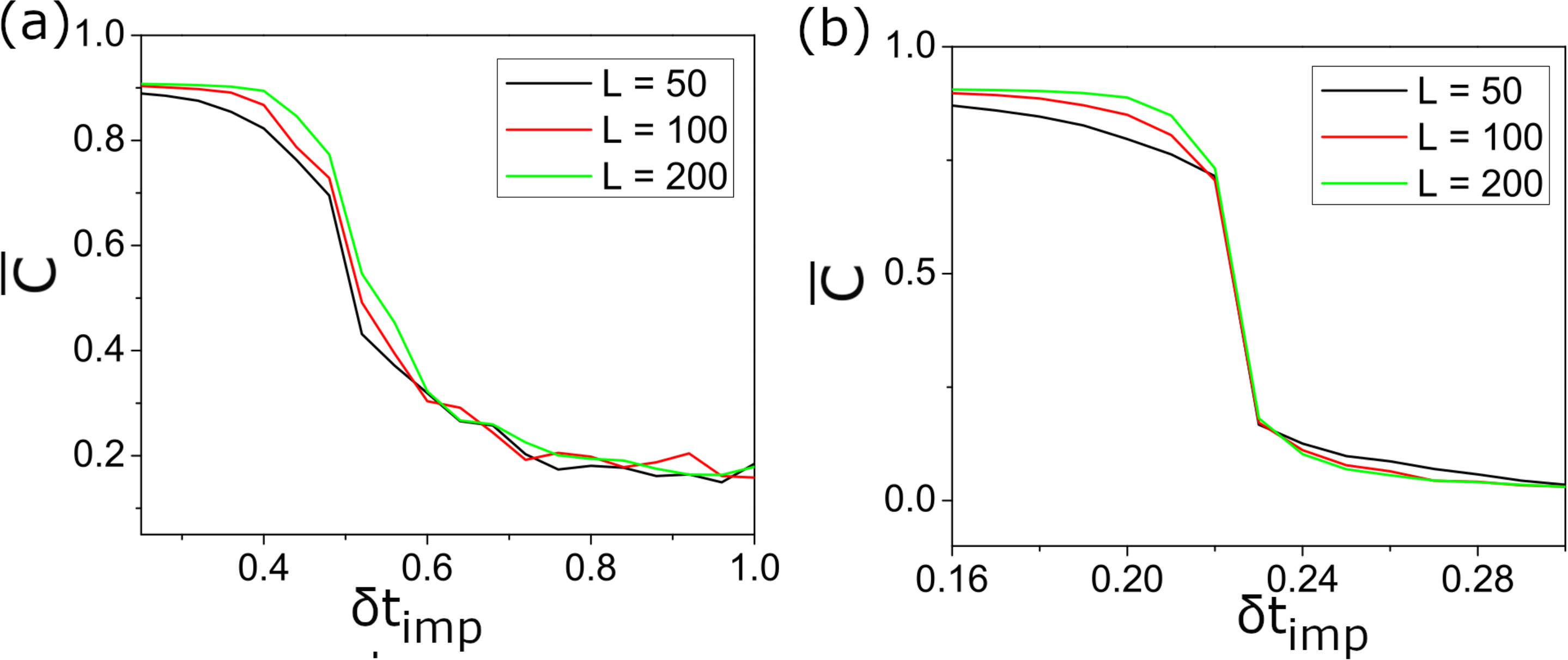}
\caption{Details of the impurity-induced crossover region in Fig.~\ref{fig:SSH_many_imp_result} (b), where we fix the impurity density at (a) $n_{imp}=40\%$ and (b) $n_{imp}=90\%$, and plot the average marker $\overline{\cal C}$ as a function of impurity strength $\delta t_{imp}$. One sees that the $n_{imp}=40\%$ has a much broader crossover region than $n_{imp}=90\%$, and moreover the width of the crossover region remains unchanged at large enough system size $L$. } 
\label{fig:crossover_SSH}
\end{center}
\end{figure}

\begin{figure*}[ht]
\begin{center}
\includegraphics[clip=true,width=1.7\columnwidth]{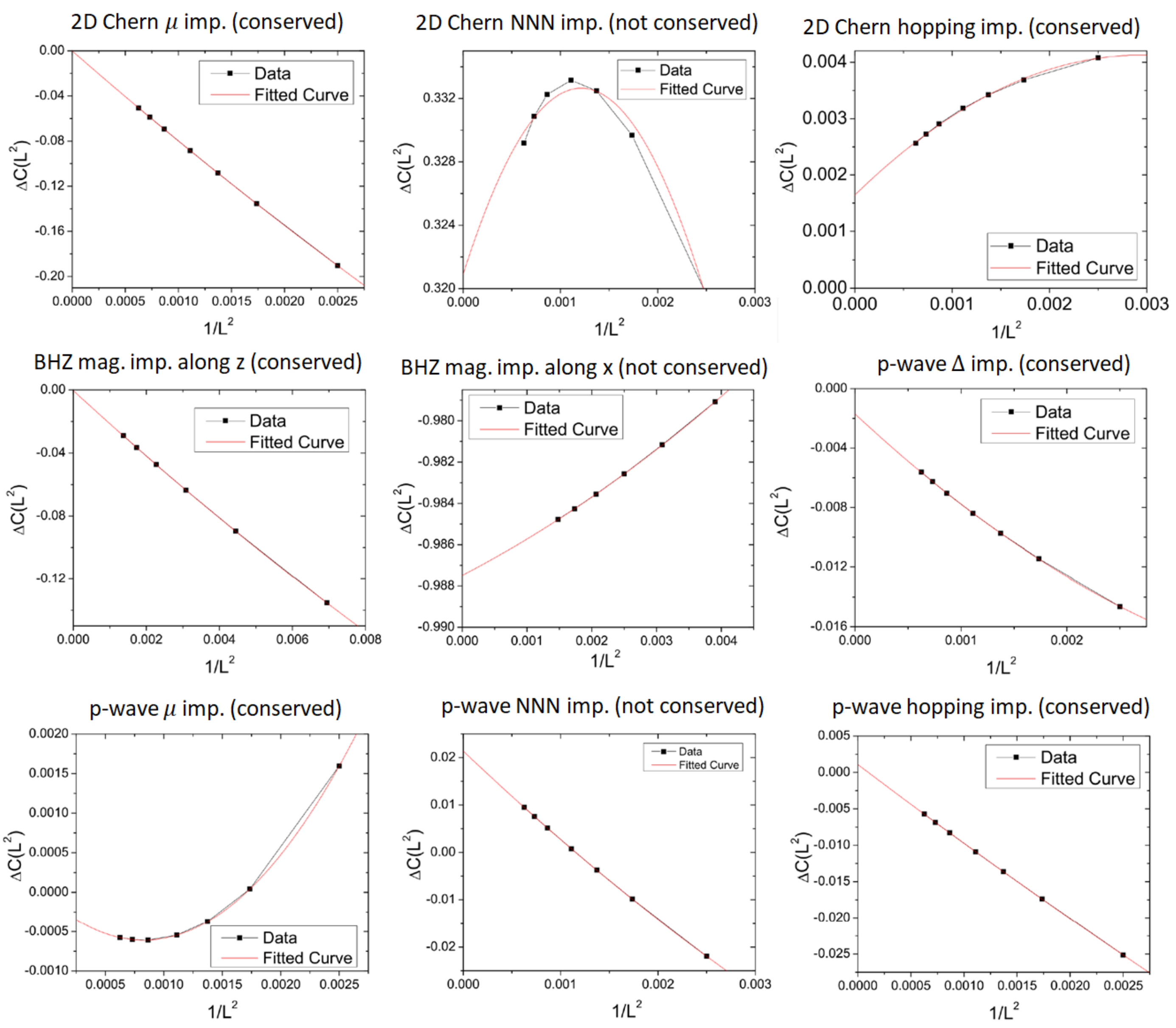}
\caption{Finite size scaling analysis of the single impurity contribution to the topological marker, $\Delta{\cal C}(L^{2})$ versus inverse system size $1/L^{2}$, for all the different kinds of impurities in all the 2D models we have investigated. At large system size $1/L^{2}\rightarrow 0$, if $\Delta{\cal C}(L^{2})\rightarrow 0$ then the spatially averaged marker is conserved in the presence of the impurity, and if $\Delta{\cal C}(L^{2})\neq 0$ then the average marker is not conserved. Here the abbreviations are imp $=$ impurity, mag $=$ magnetic, NNN $=$ next-nearest-neighbor hopping impurity, $\Delta=$ pairing impurity, and $\mu=$ potential impurity. } 
\label{fig:finite_size_scaling_figure}
\end{center}
\end{figure*}

\section{Conclusions \label{sec:conclusions}}

In summary, we investigate the robustness of topological order under the influence of disorder in a variety of theoretical models in 1D and 2D. Our survey is based on the formalism of a universal topological marker that maps the topological invariant on lattice sites. Improved by the exponentiated position operators that cure the anomaly at the boundary of the lattice, the topological marker allows to investigate the influence of any kind of impurities within tight-binding models in any dimension and symmetry class. The main issue we aim to address is whether there exists some simple principle that dictates which kinds of impurities alter the spatially-averaged marker $\overline{\cal C}$ and which kinds do not, in a way analogous to Anderson's theorem that states dilute nonmagnetic impurities do not alter the global superconductivity in $s$-wave superconductors.

We discover that in 1D and 2D, it can be proved analytically that the average marker $\overline{\cal C}$ remains quantized in the presence of dilute impurities that correspond to altering nonzero matrix elements of the lattice Hamiltonian, regardless they are potential scatterers, hopping impurities, or pairing disorder, and whether they violate the nonspatial symmetry of the host material, as supported by our numerical calculation. However, if the strength of the impurities exceeds the bulk gap and the distance between them becomes shorter than the correlation length, then the average marker $\overline{\cal C}$ no longer remains quantized. In fact, one can even use dense impurities to smoothly drive the system between topologically trivial $\overline{\cal C}=0$ and nontrivial $\overline{\cal C}=1$ phases, mimicking a first-order TPT. On the other hand, for impurities that correspond to varying zero matrix elements of the lattice Hamiltonian, we find that in general $\overline{\cal C}$ does not remain quantized.



Our results that combine analytical and numerical calculations thus provide a thorough understanding on the robustness of topological order against disorder, and moreover demonstrates the ubiquity of the universal topological marker on addressing any kind of disorder in any dimension and symmetry class. We anticipate that the marker may be used o address the local variation and global average of topological order in the presence of other kinds of inhomogeneity, such as grain boundaries and junctions, and new effects of the inhomogeneity may be discovered, which await to be explored. 



\appendix

\section{Detail of the impurity-induced crossover region in SSH model \label{apx:crossover_SSH}}

In this section, we detail the nature of the crossover region induced by many impurities in the SSH model shown in Fig.~\ref{fig:SSH_many_imp_result} (b) and (c). Firstly, we emphasize again that, by construction, a transition from the $\overline{\cal C}=1$ to the $\overline{\cal C}=0$ phase must occur at high impurity density $n_{imp}$ and large impurity strength $\delta t_{imp}$. In particular, in the maximal impurity density limit $n_{imp}=100\%$ that is practically homogeneous, the transition happens at $\delta t_{imp}=-\delta t$ since $\delta t_{imp}$ just adds to the unperturbed parameter $\delta t$ such that they together reach the critical point $\delta t+\delta t_{imp}=0$. Moreover, in this homogeneous limit $n_{imp}=100\%$, it is well-known that the transition is of second order, i.e., the average marker should jump discretely from $\overline{\cal C}=1$ to $\overline{\cal C}=0$ at the critical point $\delta t_{imp}=-\delta t$. Thus as the impurity density gradually increases to maximum $n_{imp}\rightarrow 100\%$, the crossover region is expected to narrow such that it recovers the discrete jump at $n_{imp}=100\%$. This is indeed consistent with our numerical result shown in Fig.~\ref{fig:crossover_SSH}, where $n_{imp}=90\%$ has a much narrower crossover region than $n_{imp}=40\%$. Finally, we have also verified that the finite width of the crossover region at moderate values of $n_{imp}$ remains at large system size $L$, signifying that this impurity-induced crossover region is a robust feature in the thermodynamic limit.

\section{Finite Size Scaling Analysis for Topological Marker \label{apx:finite_size_scaling}}

In this section, we conducted an analysis of the influence of finite size effects on the local marker for 2D systems, in an attempt to verify whether the average marker remains conserved in the presence of impurities. We abstain from conducting a finite size analysis on 1D systems, as the thermodynamic limit can be easily reached numerically in 1D. In contrast, such an analysis in 2D is necessary because the finite size effect renders the conservation of average marker difficult to identify at small system sizes. For this purpose, we performed a scaling analysis to determine the convergence of the difference between the total topological marker for the homogeneous and single impurity cases at a given lattice size $L^D$, given by
\begin{equation}
   \Delta C(L^2) =  \sum_{r=1}^{L^2} C(r)|_{\rm 1-imp}-\sum_{r=1}^{L^2} C(r)|_{\rm no-imp},
\end{equation}
which represents the contribution of the single impurity to the topological marker. In the thermodynamic limit $1/L^{2}\rightarrow 0$, convergence to zero $\Delta{\cal C}(L^{2})\rightarrow 0$ signifies the conservation of the average marker in the presence of the impurity. Conversely, convergence to a non-zero value $\Delta{\cal C}(L^{2})\neq 0$ suggests that the average marker is altered by the impurity. Our results are summarized in Fig.~\ref{fig:finite_size_scaling_figure}, with all the impurities and models indicated, together with whether the average marker is conserved or not. These numerical results further corroborate that our theory is correct, i.e., if the single impurity corresponds to varying a nonzero matrix element of the lattice Hamiltonian, then the average marker is conserved. For the impurities that vary the zero matrix elements of the Hamiltonian, the average marker is not conserved for all the cases we have investigated.



\bibliography{Literatur}

\end{document}